\documentclass[twocolumn,showpacs,aps,prl]{revtex4}
\usepackage{graphicx}
\usepackage{dcolumn}
\usepackage{amsmath}

\newcommand{ \rts }{ \sqrt{s_{_{\rm NN}}} }

\begin{document}
\title{\vspace*{-0.5cm}
%\begin{flushright}
%{{\small \sl version 5.a(\today) }}
%\end{flushright}
Azimuthal anisotropy of {$K_S^0$}\, and {$\Lambda + \overline{\Lambda}$}
production at mid-rapidity from Au+Au collisions at $\rts = 130$~GeV }

\author{
C.~Adler$^{11}$, Z.~Ahammed$^{23}$, C.~Allgower$^{12}$, J.~Amonett$^{14}$,
B.D.~Anderson$^{14}$, M.~Anderson$^5$, G.S.~Averichev$^{9}$, 
J.~Balewski$^{12}$, O.~Barannikova$^{9,23}$, L.S.~Barnby$^{14}$
% ...
\begin{center}(STAR Collaboration)\end{center}
}

% ...
\affiliation{$^5$University of California, Davis, California 95616}
% ...
\affiliation{$^{9}$Laboratory for High Energy (JINR), Dubna, Russia}
% ...
\affiliation{$^{11}$University of Frankfurt, Frankfurt, Germany}
\affiliation{$^{12}$Indiana University, Bloomington, Indiana 47408}
% ...
\affiliation{$^{14}$Kent State University, Kent, Ohio 44242}
% ...
\affiliation{$^{23}$Purdue University, West Lafayette, Indiana 47907}
% ...

\date{\today}
\vspace{0.5cm}
\begin{abstract}
We report STAR results on the azimuthal anisotropy
parameter $v_2$ for strange particles $K_S^0$, $\Lambda$ and
$\overline{\Lambda}$ at mid-rapidity in Au+Au collisions at $\rts =
130$~GeV at RHIC.
%-----
The value of $v_2$ as a function of transverse momentum of the
produced particles $p_t$ and collision centrality is presented for
both particles up to $p_t \sim 3.0$~GeV/c.
%-----
A strong $p_t$ dependence in $v_2$ is observed up to 2.0~GeV/c.
%-----
The $v_2$ measurement is compared with hydrodynamic model
calculations.
%-----
The physics implications of the $p_t$ integrated $v_2$ magnitude as a
function of particle mass are also discussed.
%Hydrodynamic models seem to over predict $v_2$ above $p_t
%\sim 2.0$~GeV/c.
\end{abstract}


%25.75.Wd - high nuclear energy, particle and resonance production
% 25.75.Ld - high nuclear energy, flow

\maketitle 

Measurements of azimuthal anisotropies in the transverse momentum
distribution of particles can probe early stages of ultra-relativistic
heavy-ion collisions \cite{sorge99,sorge97,ollitrault92}.
%-----
In high-energy nuclear collisions, the initial geometric anisotropy is
established from the overlap between the colliding nuclei.
%-----
The time necessary to build up this spatial anisotropy is believed to
be short because the colliding nuclei are highly Lorentz contracted in
the center-of-mass system and pass through each other at approximately
the speed of light.
%-----
During a $\sim$~5--50~fm/c period, rescattering transfers the initial
spatial anisotropy into a momentum anisotropy.
%-----
This momentum anisotropy manifests itself most strongly in the
azimuthal distribution of transverse momenta.
%-----
The extent to which the initial spatial anisotropy is transformed to
the measured momentum anisotropy depends on the initial conditions and
the dynamical evolution of the system.
%-----
In particular, anisotropy measurements for nucleus-nucleus collisions
at RHIC energies may provide information about a partonic stage that
may exist early in the collision evolution
\cite{sorge99,gyulassy99,pasi01,gyulassy01,shuryak01,zlin01}.

The transverse momentum distribution of particles can be described in
the form:
\begin{equation}
\frac{d^2N}{dp_t^2d\phi} = \frac{dN}{2\pi dp_t^2}[1 + 2\sum_n{v_n
cos(n\phi)}],
\end{equation}
where $p_t$ is the transverse momentum of the particle, $\phi$ is its
azimuthal angle with respect to the reaction
plane~\cite{yingchao96,art98} and the harmonic coefficients, $v_n$,
are anisotropy parameters.
%-----
The second coefficient $v_2$ is called {\it elliptic flow}, where flow
denotes collective behavior without necessarily implying a
hydrodynamic limit.
%-----
Recent experimental results from
RHIC~\cite{starflow1,starflow2,phenixv2qm01,phobosv2qm01} include
measurements of $v_2$ as a function of collision centrality and $p_t$
for charged particles with $p_t < 2.0$~GeV/c, and for identified
charged pions, kaons and protons for $p_t$ up to $\sim 0.8$~GeV/c.
%-----
%Since the degree of the anisotropy transfer from position to momentum
%distribution depends on the density of the system during its evolution
%and the scattering cross sections of the particles involved (parton
%and/or hadron), recent theoretical work attempted to deduce the
%initial gluon density from partonic energy loss~\cite{gyulassy01}, and
%the equation of state from hydrodynamic model
%calculations~\cite{pasi01,shuryak01}.
%-----
The degree of the anisotropy transfer from position to momentum
distribution depends on the density of the system during its evolution
and the scattering cross sections of the particles involved (parton
and/or hadron).
%-----
As a result, recent theoretical work attempted to
deduce the initial gluon density from partonic energy
loss~\cite{gyulassy01}, and the equation of state from hydrodynamic
model calculations~\cite{pasi01,shuryak01}.

Most of the anisotropic flow parameters measured to date are for
non-strange particles
\cite{starflow1,starflow2,ritter,herrmann,e87794,eos96,na4998}.
%-----
Of the studies for identified strange
particles~\cite{starflow2,4pi95,kaos98,eos98,e89501,e877la01,wa9897}
most have been at much lower collision energies.
%-----
Moreover, previous measurements of strange particle flow correspond to
directed flow, {\it i.e.} the coefficient $v_1$.
%-----
At the CERN SPS, quantitative differences between multi-strange
baryons and non-strange hadrons were observed in transverse radial
flow in Pb + Pb collisions at $\rts$ = 17~GeV~\cite{na4497,wa9798}.
%-----
A physical scenario in which multi-strange baryons do not participate
in a common expansion and thus decouple early from the collision
system due to their small hadronic cross sections, was proposed to
explain this observation~\cite{hecke98}.
%-----
This explanation suggests that it may be possible to obtain insight
into very early stages of the collisions by studying the elliptic flow
of strange particles.

In this paper, we report the first measurement of the azimuthal
anisotropy parameter~$v_2$ for the strange particles $K_S^0$,
$\Lambda$ and $\overline{\Lambda}$ from Au + Au collisions at
$\rts$~=~130~GeV.
%-----
Our measurement of $v_2$ for different centralities as a function of
$p_t$ using the Solenoidal Tracker At RHIC (STAR) extends to a $p_t$
of about 3.0~GeV/c, much higher than previously measured for
identified charged pions, kaons and protons~\cite{starflow2}.

The STAR detector~\cite{STAR}, with its azimuthal symmetry and large
acceptance, is ideally suited to measure elliptic flow.
%-----
The detector consists of several sub-systems in a large solenoidal
magnet.
%-----
For collisions in its center, the Time Projection Chamber (TPC)
measures charged tracks in the pseudo-rapidity range~$|\eta| < 1.5$
with $2\pi$ azimuthal coverage.
%-----
During the year 2000 data taking the STAR magnet operated at a
0.25~Tesla field, allowing tracking of particles with $p_t >
0.075$~GeV/c.
%-----
A scintillator barrel surrounding the TPC, the Central Trigger Barrel
(CTB), measures the charged particle multiplicity (for triggering)
from within $|\eta| < 1$.
%-----
Two zero-degree calorimeters~\cite{ZDC} located at $\pm 18.25$~m from
the nominal interaction region, sub-tending an angle~$\theta <
0.002$~radians, primarily detect fragmentation neutrons.
%-----
Two ZDCs in coincidence provide a minimum-bias trigger and the CTB is
used for a central trigger.
%-----
This analysis uses $201 \times 10^3$ minimum-bias and $180 \times
10^3$ central events.

%--======================= mass plots
%\vspace{-1.0cm}
\begin{figure}[hbtp]
%\centering\mbox{ \includegraphics[width=0.75\textwidth]{FIG_F_1.PS}}
\centering\mbox{ \includegraphics[width=0.5\textwidth]{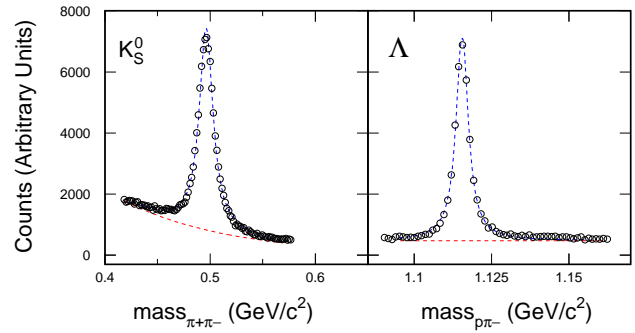}}
%\vspace{-1.5cm}
\caption{Invariant mass distributions for $\pi^{+}\pi^{-}$ showing a
$K_S^0$ mass peak (left panel) and for $p\pi^{-}$ showing a $\Lambda$
mass peak (right panel).
%-----
Fitting results are shown as dashed lines in the figure.
%-----
For presentation a greater number of events has been used for the
$\Lambda$ plot. }
\label{fig1}
\end{figure}

We reconstructed both $K_S^0 \rightarrow \pi^+ + \pi^-$ and $\Lambda
(\overline{\Lambda}) \rightarrow p + \pi^- (\overline{p} + \pi^+)$
from their charged daughter tracks detected in the STAR
TPC~\cite{tpc}.
%-----
Using the energy loss of the charged tracks in the TPC gas, we select
candidates for protons, anti-protons and pions.
%-----
The mass and the kinematic properties of the neutral particle
candidates are extracted from the decay vertex and daughter particle
kinematics.
%-----
Fig.~\ref{fig1} shows the invariant mass distributions for
$\pi^{+}\pi^{-}$ showing a $K_S^0$ mass peak and for $p\pi^-$ showing
a $\Lambda$ mass peak.  The dashed lines are fits to the background
and the signal.
%-----
We determined that the background is dominated by combinatorial counts
by rotating all positive tracks 180 degrees in the transverse plane
and reconstructing the $K^0_S$ and $\Lambda(\overline{\Lambda})$ decay
vertices.
%-----
This procedure destroys all real vertices in the TPC acceptance so
that we can describe the combinatorial contribution to the invariant
mass distributions.
%-----
The observed masses, $496 \pm 8$~MeV/$c^2$ for $\pi^+\pi^-$
and $1116 \pm 4$~MeV/$c^2$ for $p\pi$, are consistent with
accepted values~\cite{pdg} and the widths are determined by the
momentum resolution of the detector.
%-----
The particles used for the $v_2$ analysis are from the kinematic
region of $|y| \le 1.0$ and $0.2 \le p_t \le 3.2$~GeV/c for $K^0_S$ or
$0.3 \le p_t \le 3.2$~GeV/c for $\Lambda + \overline{\Lambda}$, where
$y$ is the particle's rapidity.
%-----
No significant differences in elliptic flow are observed between
$\Lambda$ and $\overline{\Lambda}$, so because of the limited
statistics, $\Lambda$ and $\overline{\Lambda}$ are summed together.

We choose the requirements for $K_S^0$ and
$\Lambda(\overline{\Lambda})$ daughter candidates to maximize
statistics.
%-----
For $K^0_S$, we require the daughter tracks to have a
distance-of-closest-approach~(dca) to the collision vertex~$>$~1.0~cm.
%-----
For the $\Lambda(\overline{\Lambda})$ reconstruction, we choose pion-like
tracks with a dca~$>$~1.5~cm and proton-like tracks with a
dca~$>$~0.8~cm.
%-----
We use the peak in the invariant mass distribution to measure the
yield of $K^0_S$ or $\Lambda + \overline{\Lambda}$ particles for
different values of $\phi$ and $p_t$.
%-----
Using the $\phi$ bin center for the value of $\phi$, we evaluate $v_2$
as a function of $p_t$ by calculating $\langle cos(2\phi) \rangle$ for
different values of $p_t$.
%-----
Using the yield to calculate $v_2 = \langle cos(2\phi) \rangle$
enables us to measure elliptic flow for identified particles beyond
the $p_t$ region where the identification of particles via their
energy loss in the TPC gas fails~\cite{starflow2}.

The real reaction plane is not known, but the event plane, an
experimental estimator of the true reaction plane, can be calculated
from the azimuthal distribution of tracks~\cite{starflow1}.
%-----
To determine the event plane, we select charged particle tracks with
at least 15 measured space points, $0.1 < p_t \le 2.0$~GeV/c and
$|\eta| < 1.0$.
%----
We also require the ratio of the number of space points to the
expected maximum number of space points for each track to be greater
than 0.52, suppressing split tracks from being counted twice.
%-----
Events are required to have a primary vertex within 75~cm
longitudinally of the TPC center.
%-----
These cuts are similar to those used in Ref.~\cite{starflow1} and our
analysis is not biased by them.

To avoid possible auto-correlations, tracks used for the $K^0_S$ or
$\Lambda(\overline{\Lambda})$ reconstruction are excluded from
the set of tracks used to calculate the event plane.
%-----
Typically this is done by measuring the azimuthal angle between a
track and the event plane calculated from all other qualifying
tracks within the same event.
%-----
Then the contribution to $v_2$ from that track is calculated.
%-----
In this analysis, where $v_2$ is not calculated on a particle by
particle basis, all tracks that might be used for the reconstruction
of $K^0_S$ or $\Lambda(\overline{\Lambda})$ are excluded from the
event plane calculation.
%-----
Only tracks with a dca~$< 1.0$~cm are used in the event plane
calculation while the $K^0_S$ vertices don't include these tracks.
%-----
In the $\Lambda + \overline{\Lambda}$ analysis all proton-like
tracks are excluded from the event plane calculation.
%In the $\Lambda + \overline{\Lambda}$ analysis, since proton-like
%tracks are likely to have a relatively small dca, they are all
%excluded from the event plane calculation.
%-----
A track is considered proton-like if its energy loss~($dE/dx$) is
within three standard deviations of that expected for protons.

When the azimuthal anisotropy is evaluated via $v_2 = \langle
cos(2\phi) \rangle$, the observed $v_2$ must be corrected to account
for the imperfect event plane resolution~\cite{starno}.
%-----
This resolution is influenced by two factors that depend on
centrality: the strength of the anisotropy signal and the number of
tracks used for the event plane calculation.
%-----
We estimate the resolution using the method of random
subevents~\cite{art98} and use the relative multiplicity, as in
Ref.~\cite{starflow1}, to measure the event centrality.
%-----
%As in Ref.~\cite{starflow1}, the multiplicity is the number of primary
%tracks that pass within 3~cm of the vertex and have $|\eta| < 0.75$.
%-----
The maximum resolution for the $K^0_S$ and $\Lambda +
\overline{\Lambda}$ analysis is found to be $0.681 \pm 0.004$ and
$0.582 \pm 0.007$ respectively and is reached in the centrality
corresponding to 25--35\% of the measured cross section.
%-----
The relatively lower resolution for the $\Lambda + \overline{\Lambda}$
analysis is caused by the exclusion of proton-like tracks from the
event plane calculation.

%=========================== centrality dependence
%\vspace{-0.5cm}
\begin{figure}[btf]
%\centering\mbox{ \includegraphics[width=0.75\textwidth]{FIG_F_2.PS}}
\centering\mbox{ \includegraphics[width=0.5\textwidth]{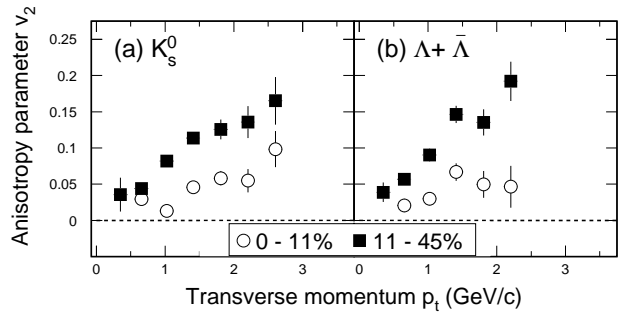}}
%\vspace{-1.0cm}
\caption{Elliptic flow~$v_2$ as a function of $p_t$ for (a) $K_S^0$
and (b) $\Lambda + \overline{\Lambda}$.
%-----
Circles and filled squares are for central (0-11\%) and mid-central
(11-45\%) collisions, respectively. }
\label{fig2}
\end{figure}

%============================= centrality dependence of v2 vs pt 
Elliptic flow as a function of transverse momentum for central and
mid-central collisions calculated from $201 \times 10^3$ minimum-bias
and $180 \times 10^3$ central events is shown in Fig.~\ref{fig2}.
%-----
The two particles show a similar $p_t$ dependence in the two
centrality intervals.
%-----
The $p_t$ dependence is stronger in more peripheral collisions than in
the central collisions.
%-----
A similar dependence was observed for charged particles in Au + Au
collisions at the same RHIC energy~\cite{starflow2}.

%--=================================== Systematic errors
For this analysis, three main sources contribute to systematic errors
in the measured anisotropy parameters: particle identification,
background subtraction, and correlations unrelated to the reaction
plane (non-flow) such as resonance decays, jets or Coulomb and
Bose-Einstein correlations~\cite{olliplb477,olliprc62}.
%-----
The contribution from the first two sources is estimated by examining
the variation in $v_2$ after changing several track and event cuts.
%-----
We estimate that these effects contribute an error of less than~$\pm$
0.005 to $v_2$.
%-----
The contribution to $v_2$ from non-flow effects, however, could be
significant, especially in peripheral collisions.
%-----
A previous study used the correlation of event plane angles from
subevents to estimate the magnitude of these
contributions~\cite{starflowqm}.
%-----
Non-flow effects are assumed to contribute to the first and second
harmonic correlations by similar amounts, so the magnitude of the
first harmonic correlation sets a limit on the non-flow contributions
to $v_2$.
%-----
That study showed that the non-flow systematic errors for charged
particles are typically +0 and -0.005, but are significantly larger in
the more peripheral events where the error increases to +0 and -0.035
for the 58--85\% most central events.
%-----
%These effects always act to increase the measured value of $v_2$ above
%its true value; so their contributions to the systematic error are
%asymmetric~\cite{starflowqm}.
%-----
These estimates are confirmed by measurements of $v_2$ using the
4th-order cumulant method, a method that is insensitive to non-flow
effects but which leads to larger statistical errors~\cite{aihong}.
%-----
%The elliptic flow measured using this method was consistent with the
%lower limit of the systematic errors from the traditional $v_2$
%analysis method.
%-----
We assume the systematic errors on $v_2$ for the neutral strange
particles $K^0_S$ and $\Lambda + \overline{\Lambda}$ are similar to
those found in the analysis of charged particles~\cite{starflow2}.
%-----
%We take the systematic errors on the 0 -- 11 \% and 11 -- 45 \% centrality
%intervals in Fig.~\ref{fig2} to be approximately

%--=============== fig v2 vs pt, minimum-bias, compared with model
%%\vspace{-1.0cm}
\begin{figure}[tbf]
%\centering\mbox{ \includegraphics[width=0.75\textwidth]{FIG_F_3.PS}}
\centering\mbox{ \includegraphics[width=0.45\textwidth]{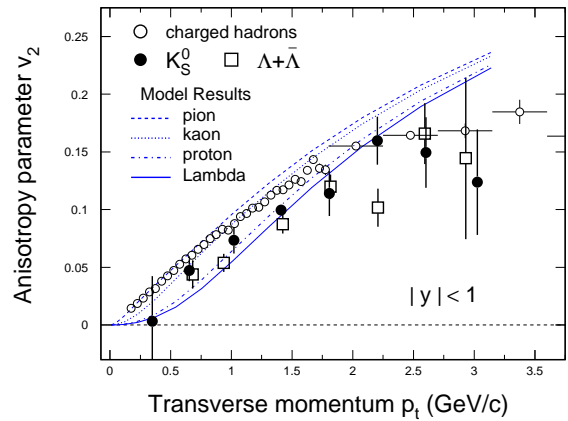}}
%\vspace{-1.0cm}
\caption{Elliptic flow~$v_2$ as a function of $p_t$ for the strange
particles $K_S^0$ (filled circle) and $\Lambda + \overline{\Lambda}$
(open squares) from minimum-bias Au+Au collisions.
%-----
For comparison, $v_2$ of charged hadrons (open circles) is also shown.
%-----
The lines are from hydrodynamic model calculations~\cite{pasi01}. }
\label{fig3}
\end{figure}

%--=========================== model comparison
To make a comparison with available hydrodynamic model calculations,
we plot $v_2(p_t)$ for both $K_S^0$ and $\Lambda + \overline{\Lambda}$
from $201 \times 10^3$ minimum-bias collisions in Fig.~\ref{fig3}.
%-----
The dashed lines represent the hydrodynamic model
calculations~\cite{pasi01} for (from top to bottom) pions, kaons,
protons, and lambdas.
%-----
Also shown in the figure is $v_2(p_t)$ for charged
hadrons~\cite{sqm2001proc}.
%-----
Within statistical uncertainty, the $K_S^0$ results are in agreement
with the $v_2$ of charged kaons~(not shown)~\cite{starflow2}.
% in the $p_t$ range they
%share ($300~\le~p_t~\le~700$~MeV/c)~\cite{starflow2}.
%In the low transverse momentum region, $p_t \le 0.8$~GeV/c, both
%values of pion and proton $v_2$ as a function of $p_t$ were well
%reproduced by the hydrodynamic calculations~\cite{starflow2,pasi01}.
We observe that $v_2$ for both strange particles increases as a
function of $p_t$ up to about 1.5~GeV/c, similar to the hydrodynamic
model prediction.
%-----
In the higher $p_t$ region however ($p_t~\ge~2$~GeV/c), the values of
$v_2$ seem to be saturated.
% and less than the hydrodynamic model
%predictions.
%-----
%This is also seen in the behavior of charged hadrons.
%-----
It has been suggested that the shape and height of $v_2$ above
2~--~3~GeV/c in a pQCD model is related to energy loss in an early,
high-parton-density, stage of the evolution~\cite{gyulassy01}.

%this saturation is related to the energy loss phenomena at high parton
%density in the early stage of the evolution~\cite{gyulassy01}.

%--=========================== v2 vs mass
%%\vspace{-1.0cm} 
\begin{figure}[btf]
%\centering\mbox{ \includegraphics[width=0.75\textwidth]{FIG_F_4.PS}}
\centering\mbox{ \includegraphics[width=0.45\textwidth]{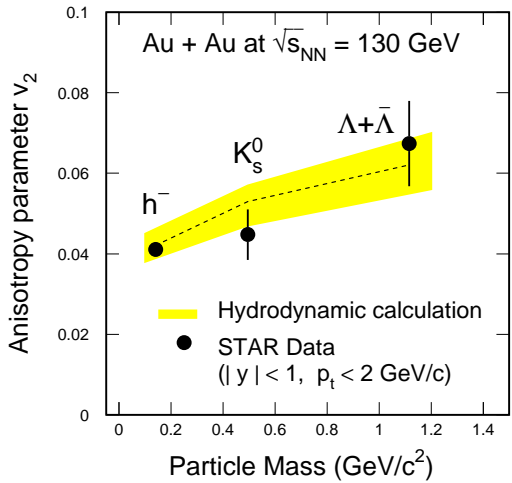}}
%\vspace{-1.5cm} 
\caption{Integrated elliptic flow $v_2$ as a function of particle
mass.
%-----
The gray-band and central line indicates the hydrodynamic model
results~\cite{pasi01}. }
\label{fig4}
\end{figure}

The $p_t$ integrated anisotropy parameters for charged hadrons,
$K_S^0$, and $\Lambda + \overline{\Lambda}$ from minimum-bias
collisions are shown in Fig.~\ref{fig4}.
%-----
The integrated values of $v_2$ are calculated by parameterizing the
yield with the inverse slope parameter of exponential fits to the
$K^0_S$ or $\Lambda + \overline{\Lambda}$ transverse mass
distributions \cite{sqm2001proc,huithesis}.
%-----
The integrated $v_2$ is dominated by the region near the particle's
mean~$p_t$ and is insensitive to the upper and lower bounds of the
integration.
%-----
Although the $v_2(p_t)$ of $\Lambda + \overline{\Lambda}$ is below the
$v_2(p_t)$ of $K_S^0$ for most $p_t$, as shown in Fig.~\ref{fig3}, the
$p_t$ integrated $v_2$ values increase with the particle mass.
%-----
This increase is partly due to the relatively higher mean~$p_t$ of the
$\Lambda + \overline{\Lambda}$ compared to the $K^0_S$.
%-----
In hydrodynamic models, although the spatial geometry of the pressure
gradient and the resultant collective velocity are the same for all
particles, massive particles tend to gain larger transverse momenta
and so develop a larger elliptic flow.
%-----
The hydrodynamic model calculations~\cite{pasi01}, shown as a
gray-band and central line, are, within errors, in agreement with this
result.
%-----
The width of the gray-band in Fig.~\ref{fig4} indicates the
uncertainties of the model calculation, mostly due to the choice of
the freeze-out conditions.
%-----
%Assumptions used in the hydrodynamic model calculations are more
%likely to be valid in the $p_t$ region near the particles mean $p_t$.
%-----
%Since this region dominates the integrated $v_2$, the apparent
%deviation of the measured $v_2$ from the hydrodynamic prediction for
%$p_t \ge 2 $~GeV/c seen in Fig.~\ref{fig3} is not strongly reflected
%in Fig.~\ref{fig4}.
The increase of $v_2$ with particle mass indicates that significant
collective motion, perhaps established early in the collision, is an
effective means to transfer geometrical anisotropy to momentum
anisotropy.
%-----
The nature of the particles during this process, however, whether
parton or hadron, and the degree of thermalization for strange
particles during the collective expansion remains an open issue.

In summary, we have reported the first measurement of the anisotropy
parameter, $v_2$, for $K_S^0$ and $\Lambda + \overline{\Lambda}$, from
Au + Au collisions at $\rts = 130$~GeV.
%Sentence 2 --------
The $v_2$ values as a function of $p_t$ from mid-central collisions
are higher at each $p_t$ than $v_2$ from central collisions.
%Sentence 3 --------
Hydrodynamic model calculations seem to adequately describe elliptic
flow of the strange particles up to a $p_t$ of 2~GeV/c.
%Sentence 4 --------
For $p_t$ above 2~GeV/c, however, the observed $v_2$ seems to saturate
whereas hydrodynamic models predict a continued increase with $p_t$.
%Sentence--------
%The $p_t$ integrated $v_2$ as a function of particle mass is
%consistent with a picture, in the hydrodynamic framework, that
%collective motion due to a pressure gradient may be an effective means
%to transfer geometrical anisotropy to momentum anisotropy.
%Sentence 5 --------
The $p_t$ integrated $v_2$ as a function of particle mass is
consistent with a hydrodynamic picture where collective motion,
established by a pressure gradient, transfers geometrical anisotropy
to momentum anisotropy.
%Sentence --------
%Although the hadronic scattering cross sections of strange and
%non-strange particles may be different, we have not yet seen any
%disagreement--for either the strange particles ($K_S^0$ and $\Lambda +
%\overline{\Lambda}$) or the non-strange particles ($\pi$ and
%p)--between the experimentally measured $v_2$ and the expected mass
%dependence from hydrodynamic model calculations.
%Sentence 6 --------
Although the hadronic scattering cross sections of strange and
non-strange particles may be different, we have yet to see deviations
in the measured $v_2$ from hydrodynamic calculations at low $p_t$ for
strange or non-strange particles.
%Sentence--------
%If indeed the measured $v_2$ proves to be independent of the hadronic
%scattering cross sections, serious consideration should be given to
%the possibility that the elliptic flow may be established before the
%hadronization epoch in a partonic phase when the hadronic cross
%sections for the final hadrons are less relevant.
%Sentence 7 --------
In a possible partonic phase prior to the hadronic epoch, the hadronic
scattering cross sections for the final hadrons are not relevant.
%Sentence 8 --------
As such, if the elliptic flow of identified particles proves to be
independent of their relative hadronic cross sections, it may be
evidence that $v_2$ is established during a partonic phase.
%Sentence 9 --------
%More theoretical work is necessary to understand these measurements
%and what they tell us about the nature of matter in the early stage of
%nuclear collisions.

%These measurements open a new direction for understanding the nature
%of matter in the early stage of nuclear collisions and call for much
%theoretical work in order to be understood.

\vspace{0.25cm} {\bf Acknowledgments:} We thank P. Huovinen for
providing the results of the hydrodynamic model calculations. We wish
to thank the RHIC Operations Group and the RHIC Computing Facility at
Brookhaven National Laboratory, and the National Energy Research
Scientific Computing Center at Lawrence Berkeley National Laboratory
for their support. This work was supported by the Division of Nuclear
Physics and the Division of High Energy Physics of the Office of
Science of the U.S. Department of Energy, the United States National
Science Foundation, the Bundesministerium fuer Bildung und Forschung
of Germany, the Institut National de la Physique Nucleaire et de la
Physique des Particules of France, the United Kingdom Engineering and
Physical Sciences Research Council, Fundacao de Amparo a Pesquisa do
Estado de Sao Paulo, Brazil, the Russian Ministry of Science and
Technology and the Ministry of Education of China and the National
Natural Science Foundation of China.

%We wish
%to thank the RHIC Operations Group at Brookhaven National Laboratory
%for their tremendous support and for providing collisions for the
%experiment. This work was supported by the Division of Nuclear Physics
%and the Division of High Energy Physics of the Office Science of the
%U.S. Department of Energy, the United States National Science
%Foundation, the Bundesministerium f\"{u}r Bildung und Forschung of
%Germany, the Institut National de la Physique Nucleaire et de la
%Physique des Particules of France, the United Kingdom Engineering and
%Physical Sciences Research Council, and the Russian Ministry of
%Science and Technology.
%-------------------------------------------------------------------

\end{document}